\documentclass[nojss]{jss}
\usepackage{amsmath,latexsym,psfrag,epsf,epsfig,amssymb,rotating}

\newcommand{\bftheta}{\hbox{\boldmath$\theta$}}

\newcommand{\bfSigma}{\hbox{\boldmath$\Sigma$}}

\newcommand{\bfmu}{\hbox{\boldmath$\mu$}}

\newcommand{\bfy}{\hbox{\boldmath$y$}}

\author{Alberto Caimo\\University of Lugano \And 
        Nial Friel\\University College Dublin}
\title{\pkg{Bergm}: Bayesian Exponential Random Graphs in \proglang{R}}

\Plainauthor{Alberto Caimo, Nial Friel} 
\Plaintitle{Bergm: Bayesian Exponential Random Graphs in R} 

\Abstract{
In this paper we describe the main featuress of the \pkg{Bergm} package for the open-source \proglang{R} software which provides a 
comprehensive framework for Bayesian analysis for exponential random graph models:  tools for parameter estimation, model 
selection and goodness-of-fit diagnostics. We illustrate the capabilities of this package describing the algorithms through a tutorial analysis of three network datasets.
}
\Plainkeywords{exponential random graph models, Bayesian inference, Bayesian model selection, Markov chain Monte Carlo}

\Address{
  Alberto Caimo\\
  Faculty of Economics\\
  University of Lugano, Switzerland\\
  E-mail: \email{acaimo.stats@gmail.com}\\
  URL: \url{https://sites.google.com/site/albertocaimo}
}

\ifpdf
    \graphicspath{{pictures/}}
\fi

\begin{document}


\section[Introduction]{Introduction}

Interest in statistical network analysis has grown massively in recent decades and its perspective and methods are now widely used in many scientific areas which involve the study of various types of networks for representing structure in many complex relational systems such as social relationships, information flows, protein interactions, etc.

Social network theory is based on the study of social relations between actors so as to understand the formation of social structures by the analysis of basic local relations. Statistical models have started to play an increasingly important role because they give the possibility to explain the complexity of social behaviour and to investigate issues on how the global features of an observed network may be related to local network structures. The observed network is assumed to be generated by local social processes which depend on the self-organizing dyadic relations between actors. The crucial challenge for statistical models in social network theory is to capture and describe the dependency giving rise to network global topology allowing inference about whether certain local structures are more common than expected.

Exponential random graph models (ERGMs) (\cite{fra:str86}; \cite{was:pat96}; \cite{rob:pat:kal:lus07}) are one of the most important family of models conceived to capture the complex dependence structure of an observed network allowing a reasonable interpretation of the underlying process which is supposed to have produced these structural properties. The dependence hypothesis at the basis of these models is that the connections between actors (edges) self-organize into small structures called configurations or network statistics. These are classical graph-theoretic structures such as degrees, cycles, etc. which can be directly incorporated in ERGMs as sufficient statistics with corresponding parameters measuring their importance in the observed network. The computational intractability of these models is the main barrier to estimation.

The Bayesian approaches for exponential random graph models developed by \cite{cai:fri11} and \cite{cai:fri13} represent one of the first complete Bayesian frameworks for these models. The doubly-intractable posterior is estimated by the use of an approximate exchange algorithm with adaptive direction sampling. This approach has proven to be effective to improve mixing and local moves on the typically thin and high posterior density region. In fact fast convergence occurs even when the algorithm starts from degenerate parameter values. This approach exhibits better performance and convergence properties with respect to classical non-Bayesian methods.

The recent progress made in the framework of network analysis has been possible thanks to the development and implementation of software able to perform computational intensive tasks. For this reason, the development of software has always represented an essential aspect of the research activity in this area.

The \pkg{Bergm} package for \proglang{R} \citep{R} implements Bayesian analysis for Exponential Random Graph Models using the methods described by \cite{cai:fri11} and \cite{cai:fri13}. The package provides a comprehensive framework for Bayesian inference and model selection using Markov chain Monte Carlo (MCMC) algorithms. It can also supply graphical Bayesian goodness-of-fit procedures that address the issue of model adequacy.
Although computationally intensive, the package is simple to use and represents an attractive way of analyzing network data as it 
offers the advantange of a complete probabilistic treatment of uncertainty.
\pkg{Bergm} is based on the \pkg{ergm} package \citep{hun:han:but:goo:mor08} which is part of the \pkg{statnet} suite of packages \citep{han:hun:but:goo:mor07} and therefore it makes use of the same model set-up and network simulation algorithms. The \pkg{ergm} and \pkg{Bergm} packages complement each other in the sense that \pkg{ergm} implements maximum likelihood-based inference whereas \pkg{Bergm} implements Bayesian inference.
The \pkg{Bergm} package has been continually improved in terms of speed performance over the 
last two years and one of the purposes of this paper is to highlight these improvements. We feel that this package now offers the
end-user a feasible option for carrying out Bayesian inference for exponential random graphs.

\begin{figure}[htp]
\centering
\includegraphics[height=9cm,keepaspectratio]{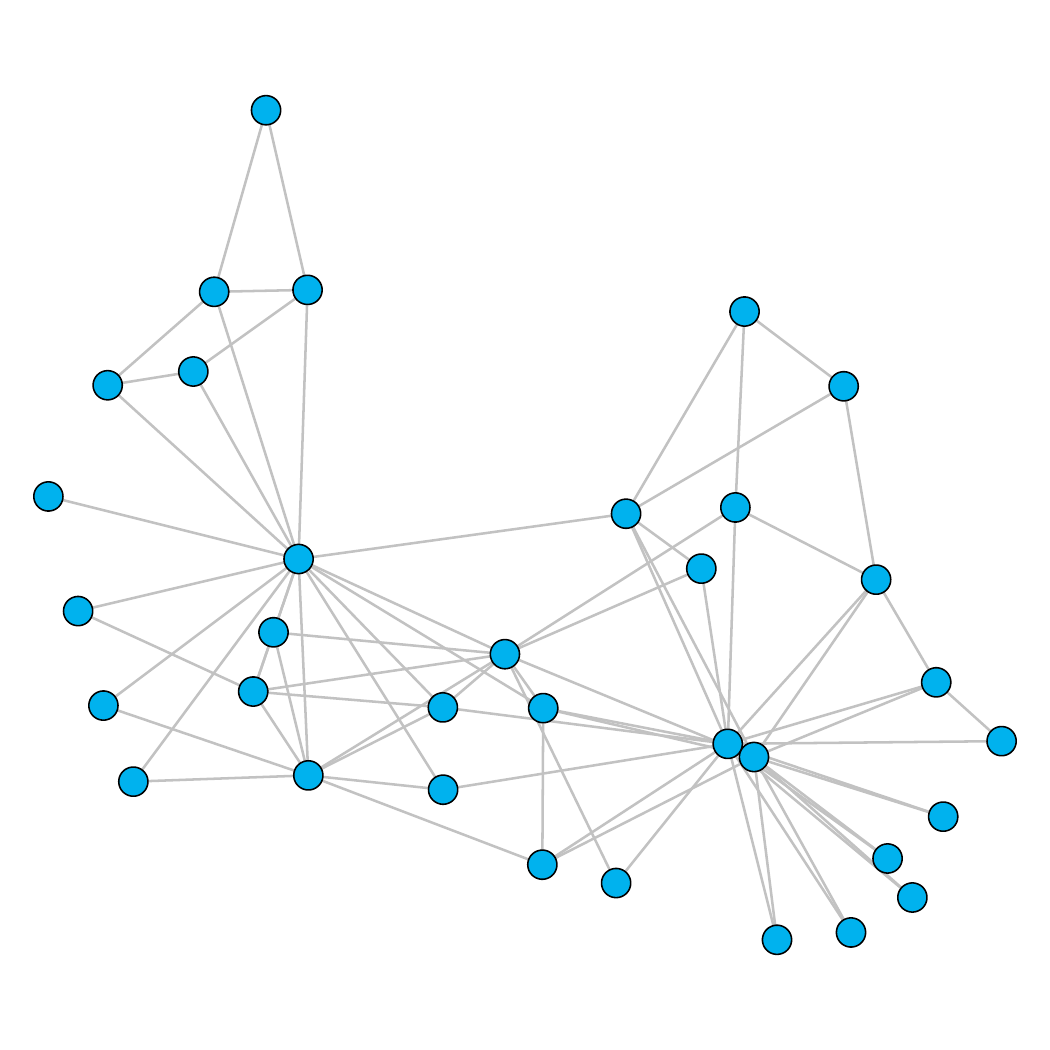}
\caption{Zachary's karate club undirected graph.}
\label{fig:graph}
\end{figure}

Three network datasets will be used throughout this tutorial for illustrative purposes: the first is the Kapferer Tailor Shop dataset \citep{kap72} whose directed edges represent work interactions in a tailor shop in Zambia (then Northern Rhodesia) and nodal attributes refer to the job status. The second network is Zachary's karate club \citep{zac77} which represents the undirected social network graph of friendships between 34 members of a karate club at a US university in the 1970s. The third is excerpt of 50 girls from the Teenage Friends and Lifestyle Study data set.
Figure~\ref{fig:graph} displays the graphs of the first two networks, Figure~\ref{fig:teen_graphs} displays the graphs of the Teenage Friends and Lifestyle Study network.
The exact \proglang{R} code used to produce these plots is given in Appendix A.

In this paper we describe how to install and load \pkg{Bergm} (Section \ref{sec:getbergm}) providing a brief summary of what Bayesian 
ERGMs are (Section \ref{sec:bbergm}). Sections \ref{sec:parbergm}, \ref{sec:bgofbergm}, and \ref{sec:bmsbergm} overview the algorithms 
and the functions used to produce posterior estimates for the parameters, Bayesian goodness-of-fit procedures and model selection 
respectively. Two examples are developed in Section~\ref{sec:kap} and Section~\ref{sec:kar}. This paper does not provide an exhaustive description of all the functionality and options available, and more 
information about the commands and methods mentioned are available through the \proglang{R} help system within the package. 
\begin{figure}[htp]
\centering
\includegraphics[height=9cm,keepaspectratio]{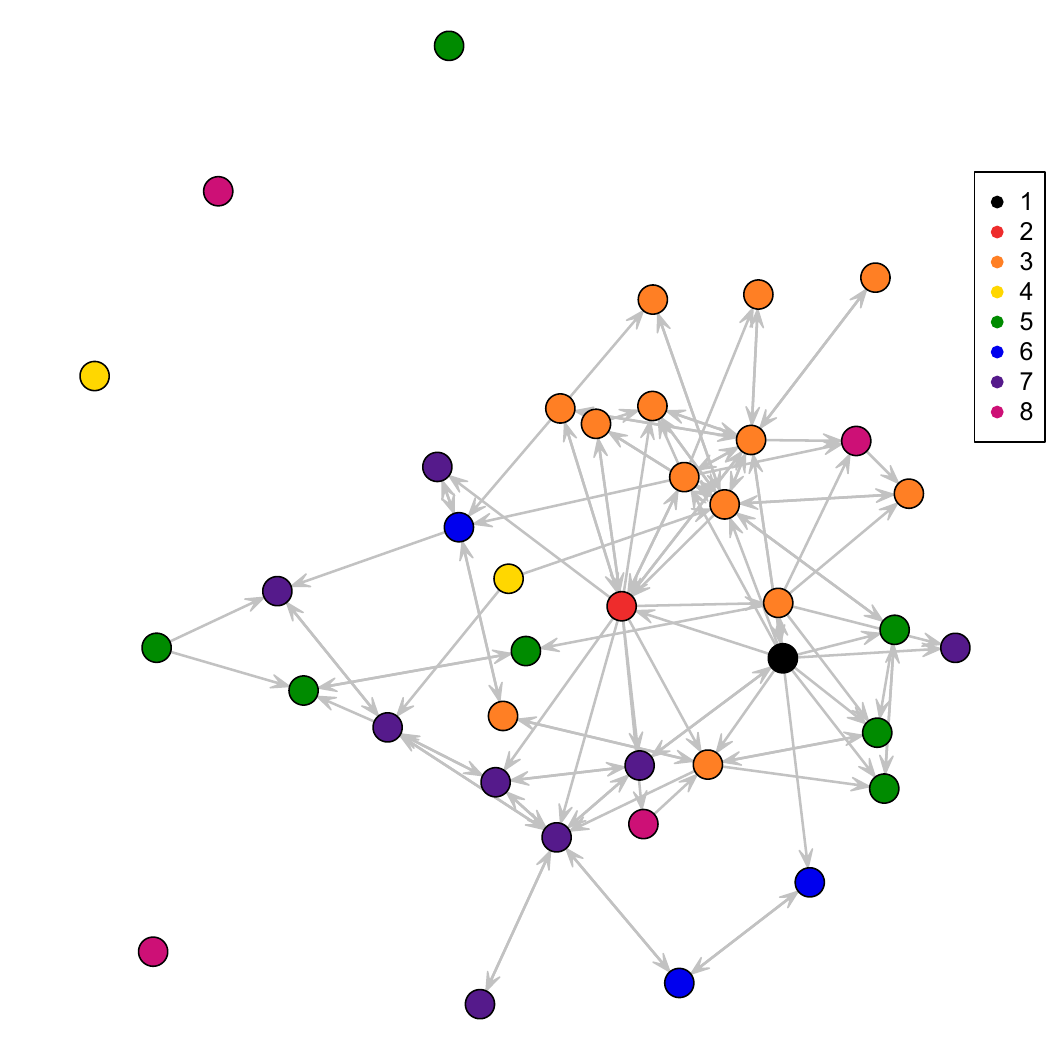}
\caption{Kapferer Tailor Shop directed graph. Nodal covariates consists of employees' occupational categories: 1=head tailor, 2=cutter, 3=line 1 tailor, 4=button machined, 5=line 3 tailor, 6=ironer, 7=cotton boy, 8=line 2 tailor.}
\label{fig:graph1}
\end{figure}

\section[Getting Bergm]{Getting \pkg{Bergm}}
\label{sec:getbergm}

The \pkg{Bergm} package can be obtained and loaded in \proglang{R} using the following commands:
\begin{CodeChunk} 
\begin{CodeInput} 
R> install.packages("Bergm")
R> library("Bergm")
\end{CodeInput}
\end{CodeChunk}
Since \pkg{Bergm} depends on \pkg{ergm} \citep{hun:han:but:goo:mor08} (which in turn depends on \pkg{network} \citep{but08}), \pkg{coda} \citep{plu:bes:cow:vin06}, and \pkg{mvtnorm} \citep{gen12}. Installing the package will automatically load all the dependencies. All of these packages are available on the Comprehensive \proglang{R} Archive Network (CRAN) at \href{http://CRAN.R-project.org}{http://CRAN.R-project.org}.

The results presented in this paper have been obtained using \proglang{R} version 2.15.1 on a Mac using \pkg{Bergm} version 2.5; \pkg{ergm} version 3.0-3; \pkg{network} version 1.7-1; \pkg{coda} version 0.15-2; and \pkg{mvtnorm} version 0.9-9993.

\section[Bayesian exponential random graphs]{Bayesian exponential random graphs}
\label{sec:bbergm}

The Bayesian approach to statistical problems is probabilistic. Inference is based on the posterior distribution which is the conditional probability of the unknown quantities given the observed ones. The posterior  distribution extracts the information in the data and provide a complete summary of the uncertainty about the unknowns.

\begin{figure}[htp]
\centering
(a)\\
\vspace{0.5cm}
\includegraphics{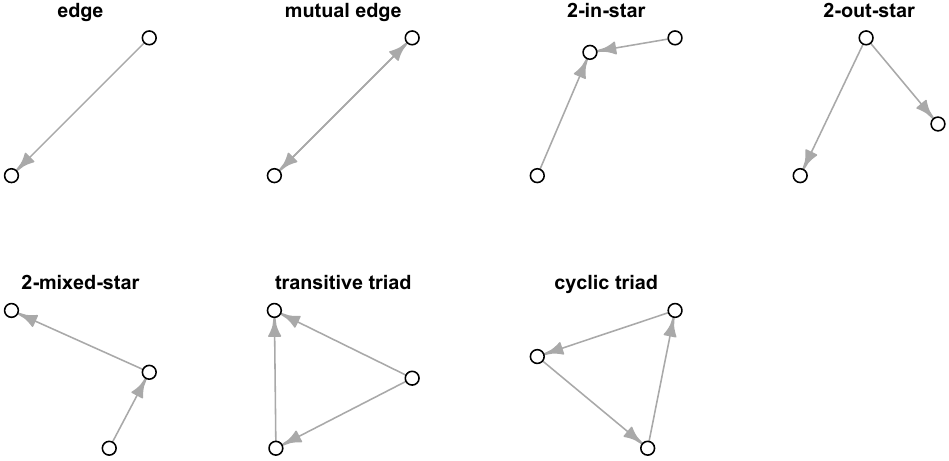}
\\ \vspace{1cm}
(b)\\
\vspace{0.5cm}
\includegraphics{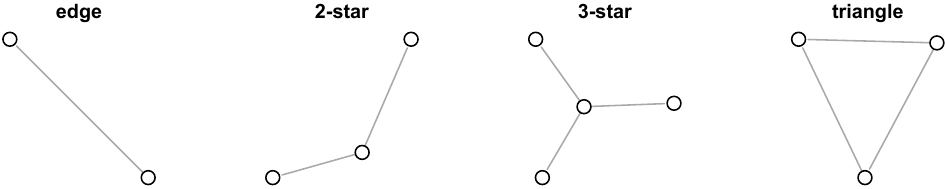}
\vspace{1cm}
\caption{Some of the most used configurations for directed (a) and undirected (b) graphs.}
\label{fig:configs}
\end{figure}

In the ERGM context (see \cite{was:pat96} and \cite{rob:pat:kal:lus07}), the purpose of Bayesian inference is to learn about the posterior distribution of the model parameters $\bftheta$ of an observed graph $\bfy$ on $n$ nodes:
\begin{equation}
p(\bftheta|\bfy) 
= \frac{p(\bfy|\bftheta)\; p(\bftheta)}{p(\bfy)}
= \frac{\exp\{\bftheta^t s(\bfy)\}}{z(\bftheta)}\;\frac{p(\bftheta)}{p(\bfy)},
\label{eqn:bergm}
\end{equation}
where $s(\bfy)$ is a known vector of sufficient network statistics (Figure~\ref{fig:configs}) \citep{mor:han:hun08}, $p(\bftheta)$ is a prior distribution placed on $\bftheta$, $z(\bftheta)$ is the likelihood normalizing constant, and $p(\bfy)$ is the model evidence.
Equation~\ref{eqn:bergm} provides a probabilistic statement about how likely parameter values are after observing the data $\bfy$. The likelihood $p(\bfy|\bftheta)$ is translated into a proper probability distribution that can be summarised by computing expected values, standard deviations, quantiles, etc.

Unfortunately the posterior distribution (Equation~\ref{eqn:bergm}) is doubly-intractable as both $z(\bftheta)$ and $p(\bfy)$ cannot be evaluated analytically \citep{kos04}. This makes the use of standard MCMC procedures infeasible.

In order to carry out Bayesian inference for ERGMs, the \pkg{Bergm} package makes use of a combination of Bayesian algorithms and 
MCMC techniques. The exchange algorithm circumvents the problem of computing the normalizing constants of the ERGM likelihoods, 
while the use of multiple chains interacting with each others (population MCMC approach) by means of adaptive direction sampling 
is able to speed up the computations and improve chain mixing quite significantly.

\section[Bayesian parameter estimation]{Bayesian parameter estimation}
\label{sec:parbergm}

In order to approximate the posterior distribution $p(\bftheta|\bfy)$, the \pkg{Bergm} package uses the exchange algorithm described in Section 4.1 of 
\cite{cai:fri11} to sample from the following distribution:
\begin{equation*}
p(\bftheta',\bfy',\bftheta | \bfy) \propto p(\bfy|\bftheta)p(\bftheta)\epsilon(\bftheta'|\bftheta) p(\bfy'|\bftheta') 
\end{equation*}
where $p(\bfy'|\bftheta')$ is the likelihood on which the simulated data $\bfy'$ are defined and belongs to the same exponential family of densities as $p(\bfy|\bftheta)$, $\epsilon(\bftheta'|\bftheta)$ is any 
arbitrary proposal distribution for the augmented variable $\bftheta'$. As we will see in the next section, this proposal 
distribution is set to be a normal centered at $\bftheta$.

At each MCMC iteration, the exchange algorithm consists of a Gibbs update of $\bftheta'$ followed by a Gibbs update of $\bfy'$, 
which is drawn from the $p(\cdot|\bftheta')$ via an MCMC algorithm \citep{hun:han:but:goo:mor08}. Then a deterministic exchange or swap 
from the current state $\bftheta$ to the proposed new parameter $\bftheta'$. This deterministic proposal is accepted with probability:
\begin{equation*}
\min\left( 1, \frac{q_{\bftheta}(\bfy')p(\bftheta') \epsilon(\bftheta|\bftheta') q_{\bftheta'}(\bfy)}
{q_{\bftheta}(\bfy)p(\bftheta) \epsilon(\bftheta'|\bftheta) q_{\bftheta'}(\bfy')} 
\times \frac{z(\bftheta)z(\bftheta')}{z(\bftheta)z(\bftheta')} \right),
\end{equation*}
where $q_{\bftheta}$ and $q_{\bftheta'}$ indicates the unnormalised likelihoods with parameter $\bftheta$ and $\bftheta'$, respectively.
Notice that all the normalising constants cancel above and below in the fraction above, in this way avoiding the need to calculate the
intractable normalising constant.

The exchange algorithm is implemented by the \code{bergm} function in the following way:
\\
\line(1,0){438}
\\
\vspace{.2cm}
{\tt for} $i = 1,\dots,N$\\
\vspace{.2cm}
\qquad {\tt 1. generate} $\bftheta'$ {\tt from} $\epsilon(\cdot|\bftheta)$\\ 
\vspace{.2cm}
\qquad {\tt 2. simulate} $\bfy'$ {\tt from} $p(\cdot|\bftheta')$\\
\vspace{.2cm}
\qquad {\tt 3. update} $\bftheta \rightarrow \bftheta'$ {\tt with the log of the probability}
\begin{equation*}
\min\left( 0,\; \left[ \bftheta - \bftheta'\right]^t \left[s(\bfy') - s(\bfy)\right]
  +\log\left[
   \frac{p(\bftheta')}
        {p(\bftheta)}\right]\right)
\end{equation*}
\vspace{.2cm}
{\tt end for}
\\
\line(1,0){438}
\\
where $s(\bfy)$ is the observed vector of network statistics and $s(\bfy')$ is the simulated vector of network statistics.

\subsection[Block-update sampler]{Block-update sampler}
\label{sec:block-update}

Step 1 of the algorithm consists in generating $\bftheta'$ from some proposal distribution within each iteration. \pkg{Bergm} uses a block-update sampler with normal proposal to simultaneously update of the parameter values in the MCMC chain: 
\begin{equation}
\epsilon(\bftheta'|\bftheta) \sim \mathcal{N}(\bftheta,\bfSigma_{\epsilon}).
\label{eqn:blockupdater}
\end{equation}
Typically, tuning the parameter $\bfSigma_{\epsilon}$ of the proposal distribution $\epsilon$ from which $\bftheta'$ is drawn represents the crucial part of the algorithm since a poor tuning of the proposal parameter $\bfSigma_{\epsilon}$ can slow down the chain's mixing rate and therefore the algorithm can take a very long time to converge to the stationary posterior density. By default $\bfSigma_{\epsilon}$ is set to a diagonal matrix with every diagonal entry equal to $0.0025$.

\subsection[Parallel adaptive direction sampler]{Parallel adaptive direction sampler}
\label{sec:ads}

\begin{figure}[htp]
\centering
\includegraphics[height=5cm,keepaspectratio]{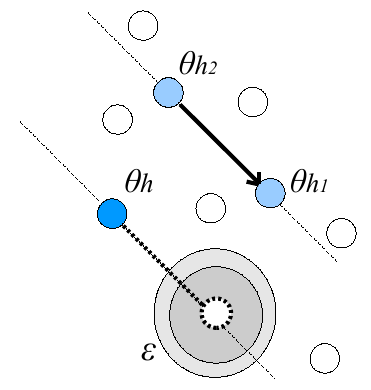}
\caption{The parallel ADS move of the current state (darker blue dot) consists of generating a new parameter value along the direction made by the difference of two randomly sampled parameter states (light blue dots) belonging to different chains plus a random term.}
\label{fig:snooker}
\end{figure}

In order to improve mixing a parallel adaptive direction sampler (ADS) \citep{gil:rob:geo94,rob:gil94} is considered: at the $i$-th iteration of the algorithm 
we have a collection of $H$ different chains interacting with one another. By construction, the state space consists of 
$\{\bftheta_1,\dots,\bftheta_H\}$ with target distribution $p(\bftheta_1|\bfy)\otimes\dots\otimes p(\bftheta_H | \bfy)$. A 
parallel ADS move consists of generating a new value $\bftheta'_h$ from the difference of two parameters $\bftheta_{h_1}$ and 
$\bftheta_{h_2}$  (randomly selected from other chains) multiplied by a scalar term $\gamma$ which is called parallel ADS move 
factor plus a random term $\epsilon$ called parallel ADS move parameter (Figure~\ref{fig:snooker}) which is equivalent to the 
block-update sampler defined in (Equation~\ref{eqn:blockupdater}). The algorithm can be summarised as follows:

\noindent \line(1,0){438}
\\
\vspace{.2cm}
{\tt for} $i = 1,\dots,N$\\
\vspace{.2cm}
\qquad {\tt for} $h = 1,\dots,H$\\
\vspace{.2cm}
\qquad\qquad {\tt 1. generate} $h_1$ {\tt and} $h_2$ {\tt such that} $h_1 \neq h_2 \neq h$\\
\vspace{.2cm}
\qquad\qquad {\tt 2. generate} $\bftheta'_h$ {\tt from} $\gamma (\bftheta_{h_1} - \bftheta_{h_2}) + \epsilon(\cdot|\bftheta_h)$\\
\vspace{.2cm}
\qquad\qquad {\tt 3. simulate} $\bfy'$ {\tt from} $p(\cdot|\bftheta'_h)$\\
\vspace{.2cm}
\qquad\qquad {\tt 4. update} $\bftheta_h \rightarrow \bftheta'_h$ {\tt with the log of the probability} \\ 
\begin{equation*}
\min\left( 0, \left[ \bftheta_h - \bftheta'_h \right]^t \left[s(\bfy') - s(\bfy)\right]
  +\log\left[ 
   \frac{p(\bftheta'_h)}
        {p(\bftheta_h)}\right]
\right)
\end{equation*}
\vspace{.2cm}
\qquad {\tt end for}\\
\vspace{.2cm}
{\tt end for}
\\
\line(1,0){438}
\\

\subsection[Kapferer tailor shop network]{Kapferer tailor shop network}
\label{sec:kap}

Consider the Kapferer Tailor Shop network and a 3-dimensional model including the following network statistics: edges (\code{edges}), mutual edges (\code{mutual}) and cyclic triples (\code{ctriple}) involving nodes with the same job status, where the job status is represented by a categorical nodal attribute variable consisting of 8 levels described in Figure~\ref{fig:graph1}.
\begin{center}
\begin{tabular}{ll}
\code{edges} & $\sum_{i\neq j}y_{ij}$\\
\code{mutual} & $\sum_{i\neq j}y_{ij}y_{ji}$ \\
\code{ctriple("job")} & $\sum_{i\neq j\neq k}y_{ij}y_{jk}y_{ki}$ where $i,j,k$ have the same job status
\end{tabular}
\end{center} 
The format of the model specification is the same of an \code{ergm} formula:
\begin{CodeChunk} 
\begin{CodeInput} 
R> formula <- y ~ edges + mutual + ctriple("job")
\end{CodeInput} 
\end{CodeChunk} 
Then we can  use the \code{bergm} function to sample from the posterior distribution using the MCMC algorithm described above. In this example we use the parallel ADS procedure described in Section~\ref{sec:ads}. By default, the number of chains in the population is set as twice the number of dimensions of the model. It is possible to choose a different number of chains by using the argument \code{nchains}. In order to perform the block-site update described in Section \ref{sec:block-update} it is necessary to set \code{nchains} = 1. For each chain, we can then set the number of burn-in iterations (\code{burn.in}) and the number of iterations after the burn-in (\code{main.iters}). The number of iterations used to simulate a network $\bfy'$ at each iteration is defined by the argument \code{aux.iters}.
\begin{CodeChunk} 
\begin{CodeInput} 
R> post.est  <- bergm(formula, 
+                     burn.in=500,
+                     gamma=0.7,
+                     main.iters=1500,
+                     aux.iters=25000)
\end{CodeInput} 
\end{CodeChunk} 
The population MCMC with parallel ADS move is the default procedure of the \code{bergm} function. The total number of iterations, e.g., the size of the posterior sample, is \code{nchains} $\times$ \code{main.iters}. The proposal covariance structure 
$\bfSigma_{\epsilon}$ is defined by the argument \code{sigma.epsilon} which is set to be a diagonal matrix with every diagonal 
entry equal to a small number. In many cases, good mixing of the chain is ensured by a sensible tuning of the parallel ADS move 
factor \code{gamma} and therefore the argument \code{sigma.epsilon} can be generally left at its default value. The parameter \code{gamma} can be easily tuned to achieve a suitable acceptance rate by starting from its default value ($0.5$). Empirically it has been observed that the value of \code{gamma} can range from $0.1$ to $1.5$ depending on the size of network and the kind of network statistics included in the model.

As said above, parallel ADS is adopted as the default procedure but it is automatically disabled in the case of uni-dimensional models where the 
block-update sampler is used and the argument \code{gamma} is used to tune the variance of the normal proposal distribution $\epsilon$.

\begin{figure}[htp]
\centering
\includegraphics[height=15cm,keepaspectratio]{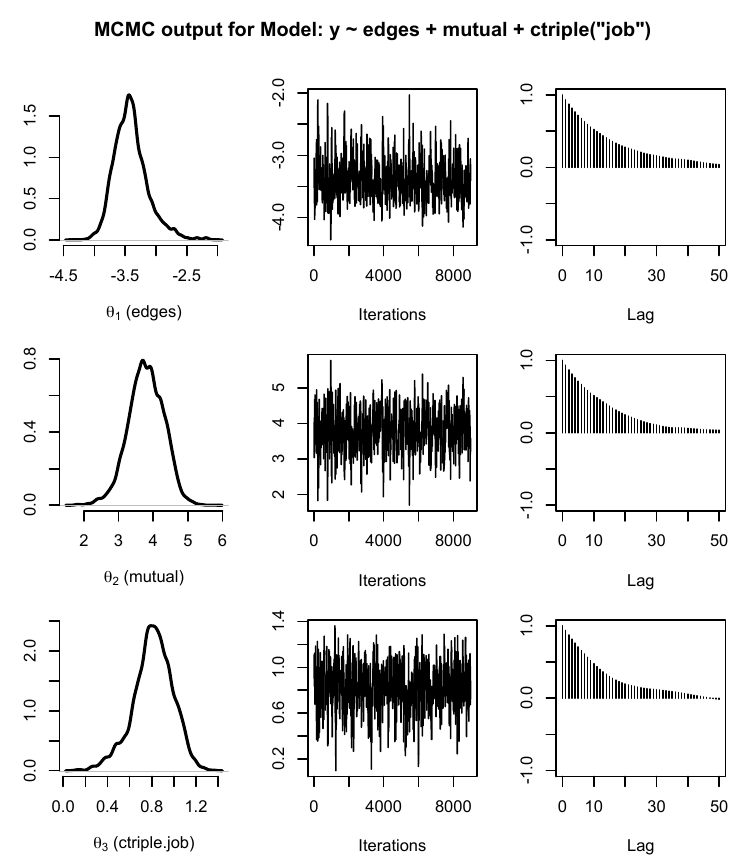}
\caption{MCMC diagnostics for the overall chain. The 3 plot columns, from left to right, are: estimated marginal posterior densities, trace plots, and autocorrelation plots.}
\label{fig:mcmcout}
\end{figure}

After completing the estimation, \code{post.est} is an object of the class \code{bergm} and contains a list of attributes among 
which are the real and CPU time (in seconds) taken by the estimation process:
\begin{CodeChunk} 
\begin{CodeInput} 
R> post.est$time
\end{CodeInput}
\begin{CodeOutput}
   user  system elapsed 
153.552   1.097 156.154 
\end{CodeOutput}
\end{CodeChunk}
It is possible to visualise the results of the MCMC estimation by using the \code{bergm.output} function which is based on the 
\pkg{coda} package \citep{plu:bes:cow:vin06} which is an \proglang{R} package for MCMC output analysis and diagnostics.
\begin{CodeChunk} 
\begin{CodeInput} 
R> bergm.output(post.est,lag.max=50)
\end{CodeInput}
\begin{CodeOutput}
 MCMC results for Model: y ~ edges + mutual + ctriple("job") 

 Posterior mean: 
          theta1 (edges) theta2 (mutual) theta3 (ctriple.job)
Chain 1       -3.4078806       3.7965993            0.8307185
Chain 2       -3.3228645       3.6495662            0.8075557
Chain 3       -3.4008949       3.8034615            0.8032079
Chain 4       -3.4220318       3.8276477            0.7884897
Chain 5       -3.4211673       3.7960881            0.8026476
Chain 6       -3.4566367       3.8762017            0.8350634

 Posterior sd: 
          theta1 (edges) theta2 (mutual) theta3 (ctriple.job)
Chain 1        0.3288498       0.5853608            0.1848051
Chain 2        0.2720809       0.4739855            0.1782770
Chain 3        0.2606994       0.4939138            0.1875215
Chain 4        0.2530538       0.4827011            0.1817614
Chain 5        0.2731412       0.4997858            0.1635572
Chain 6        0.2506724       0.4709364            0.1691319

          Acceptance rate:
Chain 1          0.2713333
Chain 2          0.2340000
Chain 3          0.2393333
Chain 4          0.2280000
Chain 5          0.2240000
Chain 6          0.2353333


 Overall posterior density estimate: 
           theta1 (edges) theta2 (mutual) theta3 (ctriple.job)
Post. mean     -3.4052460       3.7915941            0.8112805
Post. sd        0.2772942       0.5072483            0.1784213

 Overall acceptance rate: 0.238666666666667  
 
\end{CodeOutput}
\end{CodeChunk}
The output above shows the results of the MCMC estimation: posterior means and standard deviations, and acceptance rates for every 
chain in the population and for the overall chain.
Notice that the posterior summaries of each chain are consistent with each other.
Figure~\ref{fig:mcmcout} displays the MCMC diagnostic plots produced by the \code{bergm.output} function. In this example, we 
observe a low density effect expressed by the negative value of the posterior mean of the edge effect parameter ($\theta_1$) combined 
with the positive mutuality and transitivity within people having the same job status expressed by the mutual edge parameter ($\theta_2$) and cyclic triple parameter ($\theta_3$) respectively. The overall acceptance rate is around $24\%$ and the autocorrelation is negligible after lag 50.

\section[Bayesian goodness-of-fit diagnostics]{Bayesian goodness-of-fit diagnostics}
\label{sec:bgofbergm}

An important contribution of this article is to propose a Bayesian procedure to establish whether the estimated parameter posterior of the model achieves a good fit to the key topological features of the observed network.

The \code{bgof} function provides a useful tool for assessing Bayesian goodness-of-fit so as to examine the fit of the data to the posterior model obtained by the \code{bergm} function. The observed network data $\bfy$ are compared with a set of networks $\bfy_1, \bfy_2, \dots, \bfy_S$ simulated from $S$ independent realisations $\bftheta_1, \bftheta_2,\dots,\bftheta_S$ of the posterior density estimate. This comparison is made in terms of high-level characteristics $g(\cdot)$ such as higher degree distributions, etc. (see \cite{hun:goo:han08}).
The algorithm can be summarised as follows:
\\
\line(1,0){438}
\\
\vspace{.2cm}
{\tt for} $i = 1,\dots,S$\\
\vspace{.2cm}
\qquad\qquad {\tt 1. sample} $\bftheta_i$ {\tt from the estimate of} $p(\bftheta|\bfy)$\\
\vspace{.2cm}
\qquad\qquad {\tt 3. simulate} $\bfy_i$ {\tt from} $p(\cdot|\bftheta_i)$\\
\vspace{.2cm}
\qquad\qquad {\tt 4. calculate} $g(\bfy_i)$ \\ 
\vspace{.2cm}
{\tt end for}
\\
\line(1,0){438}
\\

\begin{figure}[htp]
\centering
\includegraphics[height=12cm,keepaspectratio]{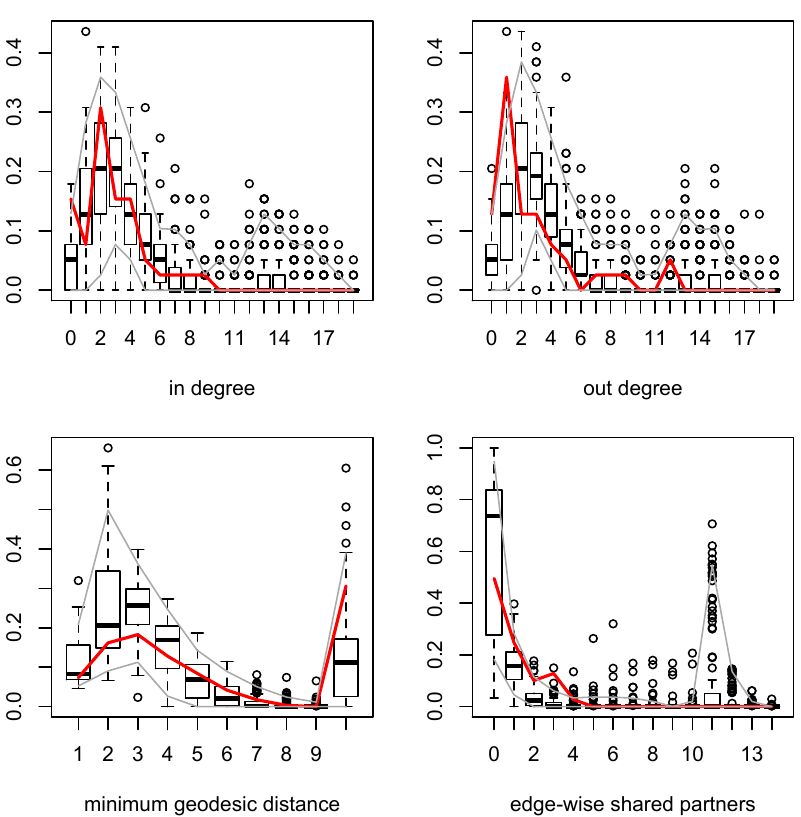}
\caption{Bayesian goodness-of-fit diagnostics. The red line displays the goodness of fit statistics for the observed data together
with boxplots of goodness of fit statistics based on $100$ simulated networks from the posterior distribution.}
\label{fig:bgofout}
\end{figure}

For example, the code below is used to compare the Kapferer Tailor Shop network with a series of networks simulated from $S=100$ 
random realisations (\code{sample.size}) of the estimated posterior distribution \code{post.est} using $50,000$ iterations 
(\code{aux.iters}) for the network simulation step. The \code{bgof} function may take a few seconds to run and, at the end of 
the execution, it will automatically plot the results as shown in Figure~\ref{fig:bgofout}.
\begin{CodeChunk} 
\begin{CodeInput} 
R> bgof(post.est,
+       sample.size=100,
+       aux.iters=50000,
+       directed=TRUE,
+       n.ideg=20,
+       n.odeg=20,
+       n.dist=10,
+       n.esp=15) 
\end{CodeInput} 
\end{CodeChunk}
The set of statistics used for the comparison of directed networks includes the in-degree distribution, the out-degree distribution, 
the minimum geodesic distance distribution and the edgewise shared partner distribution. The arguments \code{n.ideg}, \code{n.odeg}, 
\code{n.dist}, and \code{n.esp} indicates the number of boxplots to plot for each distribution respectively.

In Figure \ref{fig:bgofout} we see, based on the various goodness of fit statistics, that the networks simulated from the posterior 
distribution are in reasonable agreement with the observed network. We can therefore conclude that the data are a reasonable fit
to the model, despite its simplicity.

\section[Bayesian model selection]{Bayesian model selection}
\label{sec:bmsbergm}

An important problem in statistical analysis is the choice of an optimal model from a set of \textit{a priori} competing models. 
In the ERGM context, this task translates into the choice of which subset of network statistics should be included into the model. 

Let $m$ indicate a particular model from a set of competing models with corresponding parameters $\bftheta_m$. 
Following the Bayesian paradigm, interest focuses on exploring the posterior distribution,
\begin{equation*}
 p(m, \bftheta_m|\bfy) \propto p(\bfy|m,\bftheta_m)\; p(\bftheta_m|m) \; p(m)
\end{equation*}
where $p(\bftheta_m|m)$ and $p(m)$ are prior distributions within model $m$, and on model $m$, respectively. 
The reversible jump Markov chain Monte Carlo (RJMCMC) algorithm \citep{gre95} was designed to explore this type of posterior 
distribution across the joint model and parameter space. It is therefore a type of MCMC algorithm that allows one to jointly
explore the uncertain between and within models.

This approach is very appealing since it relies exclusively on probabilistic considerations but is very challenging from a 
computational viewpoint. As stated above, the intractability of the likelihood normalising constant $z(\bftheta_m)$ in Equation~\ref{eqn:bergm} 
renders standard RJMCMC techniques infeasible. However, the exchange algorithm used for parameter estimation can be easily generalised 
so as to include model indicators.

The auto-RJ exchange algorithm described in \cite{cai:fri13} represents a trans-dimensional RJMCMC extension of the exchange algorithm 
involving an independence sampler based on a distribution fitting a parametric density approximation to the within-model posterior. 
This approach overcomes the issue of the likelihood intractability sampling from:
\begin{equation}
p(\bftheta'_{m'},\bftheta_m, m', m, \bfy' | \bfy) 
   \propto 
   p(\bfy|\bftheta_m, m) p(\bftheta_m|m) p(m) w(\bftheta'_{m'}|m') h(m'|m) p(\bfy'|\bftheta'_{m'}, m')
\label{eq:rjtrgt}
\end{equation}
where $m$ and $m'$ are two competing models, $p(\bfy|\bftheta_m, m)$ and $p(\bfy'|\bftheta'_{m'}, m')$ are the two likelihoods for the 
data $\bfy$ under model $m$ and the simulated data $\bfy'$ under model $m'$ respectively, $p(\bftheta_m|m)$ and $p(m)$ are the priors 
for the parameter and the respective model $m$, $w(\cdot)$ is a within-model proposal (independence sampler) which fit a 
parametric density approximation to the model posteriors and $h(\cdot)$ is a between-model proposal. Notice that the marginal 
distribution for $\bftheta'_{m'}$ and $m'$ in Equation~\ref{eq:rjtrgt})  is the target distribution of interest $p(\bftheta_m,m|\bfy)$.

The \code{bergmS} function implements the auto-RJ exchange algorithm which consists of two parts: an offline step and an online step. 
In the online step, samples from the posterior $p(\bftheta_m|\bfy,m)$ are gathered from each competing model using the \code{bergm} 
function and then approximated by normal distributions $\mathcal{N}(\hat{\bfmu}_{m},\hat{\bfSigma}_{m})$ determined by the first 
and second moments from a sample from the model.

The second step (online run) of the algorithm consists of a Gibbs update of $m'$ followed by a Gibbs update of $\bftheta'_{m'}$ which 
is generated via the independence sampler $w(\bftheta'_{m'}|m') \sim \mathcal{N}(\hat{\bfmu}_{m'},\hat{\bfSigma}_{m'})$. This is 
followed by a Gibbs update of $\bfy'$ which is generated from $p(\cdot|\bftheta'_{m'},m')$. Then a deterministic exchange move from 
a current state $(\bftheta_m,m)$ to the proposed new state $(\bftheta'_{m'},m')$ is accepted with probability:
\begin{equation*}
\min \left\lbrace 
1,
   \frac{q_{\bftheta_m,m}(\bfy')}
        {q_{\bftheta_m,m}(\bfy)}
   \frac{q_{\bftheta'_{m'},m'}(\bfy)}
        {q_{\bftheta'_{m'},m'}(\bfy')}
   \frac{p(\bftheta'_{m'}|m')}
        {p(\bftheta_m|m)}
   \frac{p(m')}
        {p(m)}
   \frac{w(\bftheta_m|\hat{\bfmu}_{m},\hat{\bfSigma}_{m})}
        {w(\bftheta'_{m'}|\hat{\bfmu}_{m'},\hat{\bfSigma}_{m'})}
   \frac{h(m|m')}{h(m'|m)}
     \right\rbrace
\end{equation*}
where $q_{\bftheta_m,m}$ and $q_{\bftheta'_{m'},m'}$ indicates the unnormalised likelihoods under model $m$ with parameter $\bftheta_m$ and under model $m'$ with parameter $\bftheta'_{m'}$ respectively.

The structure of the \code{bergmS} function can be described in the following way:
\\
\line(1,0){438}
\\
\vspace{.2cm}
{\tt for} $i = 1,\dots,N$\\
\vspace{.2cm}
\qquad {\tt 1. generate} $m'$ from $h(\cdot|m)$\\
\vspace{.2cm}
\qquad {\tt 2. generate} $\bftheta'_{m'} \sim \mathcal{N}(\hat{\bfmu}_{m'},\hat{\bfSigma}_{m'})$ \\
\vspace{.2cm}
\qquad {\tt 3. simulate} $\bfy'$ {\tt from} $p(\cdot|\bftheta'_{m'},m')$\\
\vspace{.2cm}
\qquad {\tt 4. update} $(\bftheta_m,m) \rightarrow (\bftheta'_{m'},m')$ {\tt with the log of the probability:}
\begin{equation*}
\min \left(0,    
\bftheta_m^{t} 
[s_{m}(\bfy') - s_{m}(\bfy)] 
+ \bftheta_{m'}^{'t} 
[s_{m'}(\bfy)-s_{m'}(\bfy')]
  +\log\left[ 
   \frac{p(\bftheta'_{m'})}
        {p(\bftheta_m)}
        \frac{w(\bftheta_m|\hat{\bfmu}_{m},\hat{\bfSigma}_{m})}
             {w(\bftheta'_{m'}|\hat{\bfmu}_{m'},\hat{\bfSigma}_{m'})}\right]
        \right)
\end{equation*}
\vspace{.2cm}
{\tt end for}
\\
\line(1,0){438}
\\
where $s_{m}(\bfy)$ and $s_{m'}(\bfy)$ are the observed vectors of network statistics under model $m$ and $m'$ respectively, and 
$s_{m}(\bfy')$ and $s_{m'}(\bfy')$ are the simulated vector of network statistics under model $m$ and $m'$ respectively. Here, we have assumed the model priors and the between model proposals to be uniform. The reader is referred to \cite{cai:fri13} for more details on this algorithm.

\subsection[Karate club network]{Karate club network}
\label{sec:kar}

In this example, we consider the Karate club network and we propose three competing models to fit the data using a set of new specification statistics introduced by \cite{hun:han06} and \cite{hun07}: geometrically weighted edgewise shared partners (\code{gwesp}) and geometrically weighted degrees (\code{gwdegree}):
\begin{center}
\begin{tabular}{ll}
\code{gwesp} & $e^{\phi_v} \sum_{k=1}^{n-2}
\left \{ 1-\left( 1 - e^{-\phi_v} \right)^{k} \right \} EP_k(\bfy)$  \\
\code{gwdegree} & $e^{\phi_u} \sum_{k=1}^{n-1} 
\left\{ 1- \left( 1 - e^{-\phi_u} \right )^{k} \right \} D_k(\bfy)$
\end{tabular}
\end{center}
where the scale parameters $\phi_v=0.2$ and $\phi_u=0.8$. $D_k(\bfy)$ is the number of pairs that have exactly $k$ common neighbours and $EP_k(\bfy)$ is the number of connected pairs with exactly $k$ common neighbours.

The specification of these models requires the creation of a list of formulas:
\begin{CodeChunk}
\begin{CodeInput}
R> formulae <- c(y ~ edges + gwesp(0.2,fixed=TRUE),
+                y ~ edges + gwdegree(0.8,fixed=TRUE),
+                y ~ edges + gwesp(0.2,fixed=TRUE) 
+                          + gwdegree(0.8,fixed=TRUE))
\end{CodeInput}
\end{CodeChunk}
The \code{bergmS} command is then used to carry out the algorithm. To do this we have to specify several arguments for both the offline and online step.
 
The offline run consists of running the \code{bergm} function for each of the models proposed. Therefore we set some arguments \code{main.iters}, \code{burn.ins}, \code{gammas} which are vectors containing values for \code{bergm} arguments: \code{main.iters}, \code{burn.in}, \code{gamma} for each competing model. 

\begin{figure}[htp]
\centering
\includegraphics[height=5cm,keepaspectratio]{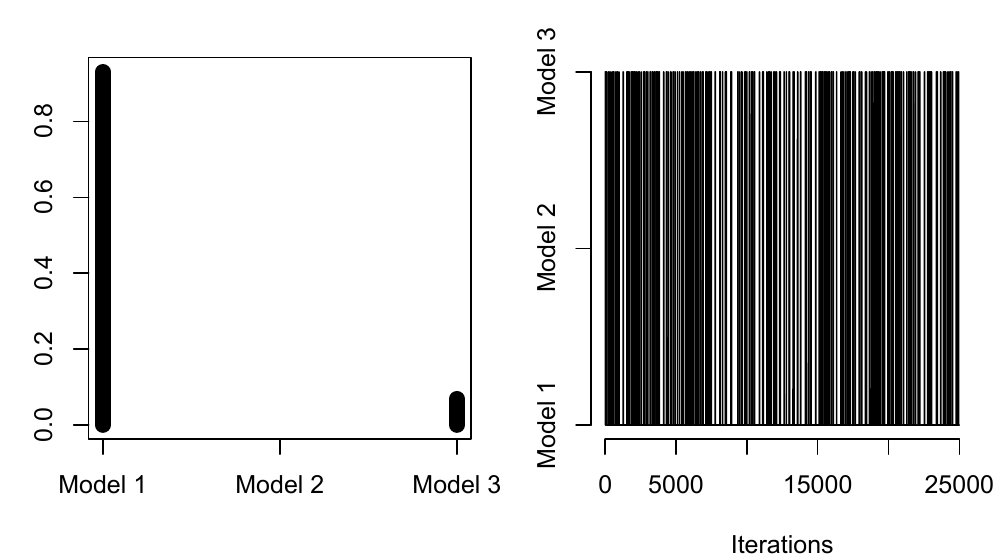}
\caption{MCMC diagnostics: posterior model probabilities.}
\label{fig:Mselect}
\vspace{1cm}
\includegraphics[height=12cm,keepaspectratio]{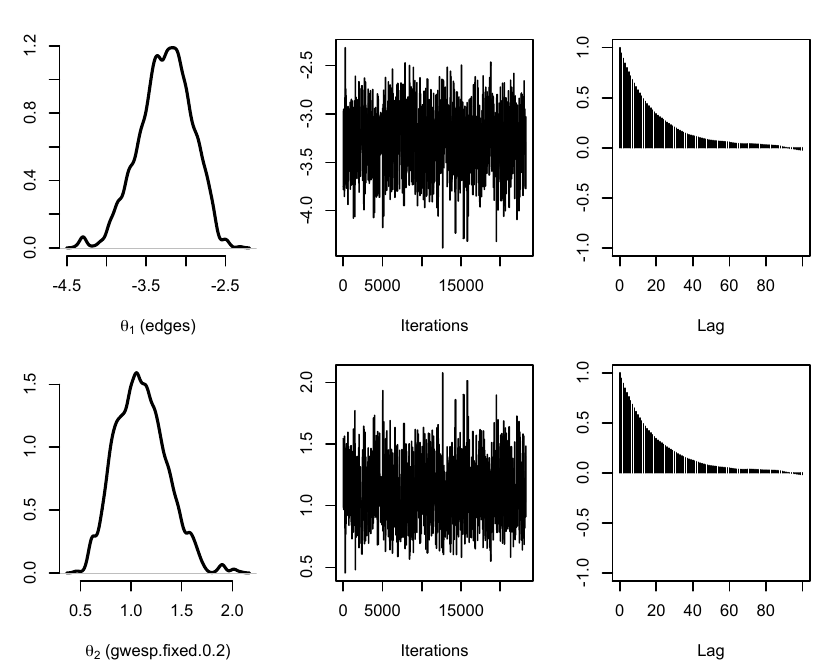}
\caption{MCMC diagnostics: posterior parameter probabilities.}
\label{fig:Pselect}
\end{figure}

The argument \code{iters} refers to the number of iterations used for the online run. The number of MCMC steps used for network 
simulation is specified as usual by the argument \code{aux.iters} and this will be used in both the offline and the online step. 
The command below should take around 10 minutes depending on the CPU speed of the computer.
\begin{CodeChunk}
\begin{CodeInput}
R> mod.sel <- bergmS(formulae,
+                    iters=25000, 
+                    aux.iters=10000,
+                    main.iters=rep(700,3),
+                    burn.ins=rep(100,3),
+                    gammas=c(1,1,0.8))
\end{CodeInput}
\end{CodeChunk}
The \code{bergmS.output} function produces the MCMC diagnostics for each competing model explored by the MCMC algorithm. Figure~\ref{fig:Mselect} and \ref{fig:Pselect} display the plots regarding the posterior model and parameter density estimate for the best model, respectively.
\begin{CodeChunk}
\begin{CodeInput}
R> best.mod <- bergmS.output(mod.sel,lag.max=100)
\end{CodeInput}
\begin{CodeOutput}
 BEST MODEL 
 ----------
Model 1: y ~ edges + gwesp(0.2, fixed = TRUE)

 Posterior parameter estimate: 
                         Post. mean:  Post. sd:
theta1 (edges)            -3.2574625  0.3278196
theta2 (gwesp.fixed.0.2)   1.1008261  0.2515162

 Within-model acceptance rate: 0.26 
 

Model 3: y ~ edges + gwesp(0.2, fixed = TRUE) + gwdegree(0.8, fixed = TRUE)

 Posterior parameter estimate: 
                         Post. mean:  Post. sd:
theta1 (edges)            -3.4916584  0.5146103
theta2 (gwesp.fixed.0.2)   1.1824831  0.2737858
theta3 (gwdegree)          0.4783638  0.5852124

 Within-model acceptance rate: 0.14 
 
BF_13 = 13.4508670520231

Between-model acceptance rate: 0.04
\end{CodeOutput}
\end{CodeChunk}
In the results above we have the posterior parameter estimates for two of the competing models (Model 2 has not been visited through the MCMC runs) with respective within-model acceptance rates and an estimate of the Bayes Factor (about $13$) for the comparison between Model 1 and Model 3 which makes clear that there is evidence that Model 1 is the best model of the set. This implies that the observed network is not enhanced by the effect captured by the geometrically weighted degree network statistic.

\begin{figure}[htp]
\centering
\includegraphics{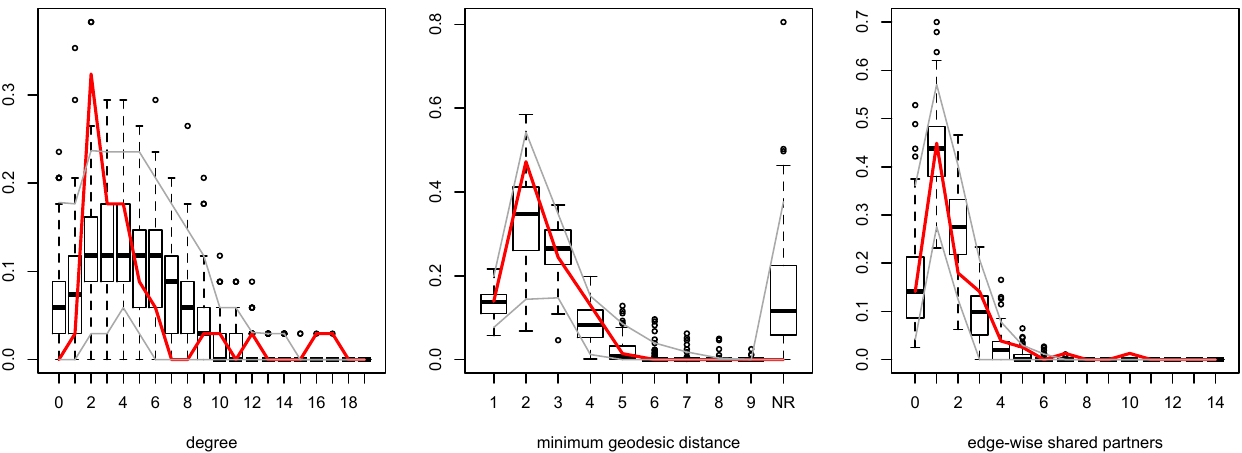}
\caption{Bayesian goodness-of-fit diagnostics.}
\label{fig:Bselect2}
\end{figure}

After running the command \code{bergmS.output}, it is possible to perform a Bayesian goodness-of-fit tests. In this case, since the observed network is undirected, the set of high-level statistics include the degree distribution in place of the in-degree and out-degree distributions.
\begin{CodeChunk}
\begin{CodeInput}
R> bgof(best.mod,
+       aux.iters=30000,
+       n.deg=20,
+       n.dist=10,
+       n.esp=15)
\end{CodeInput}
\end{CodeChunk}
In this example, the observed data appears to be a reasonable fit to the posterior distribution of the model selected (Model 1),
based on the goodness of fit geodesic and the shared partners statistics displayed in Figure~\ref{fig:Bselect2}.

\section[Example: Teenage Friends and Lifestyle Study]{Example: Teenage Friends and Lifestyle Study}
\begin{figure}[htp]
\centering
\includegraphics[height=6cm,keepaspectratio]{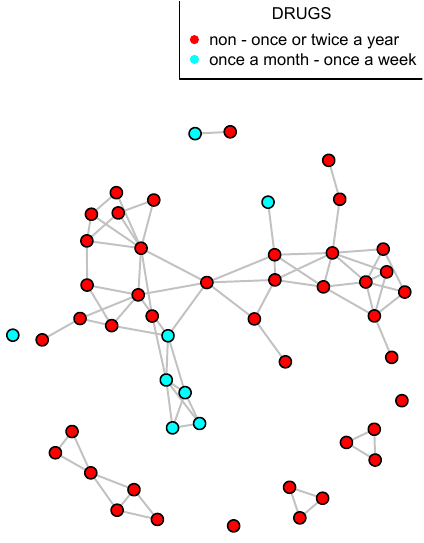}\;\;
\includegraphics[height=6cm,keepaspectratio]{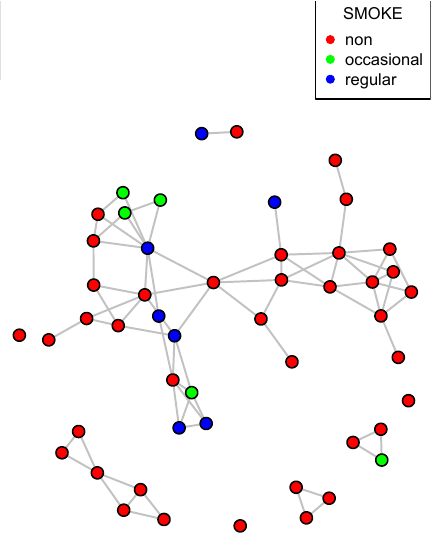}\;\;
\includegraphics[height=6cm,keepaspectratio]{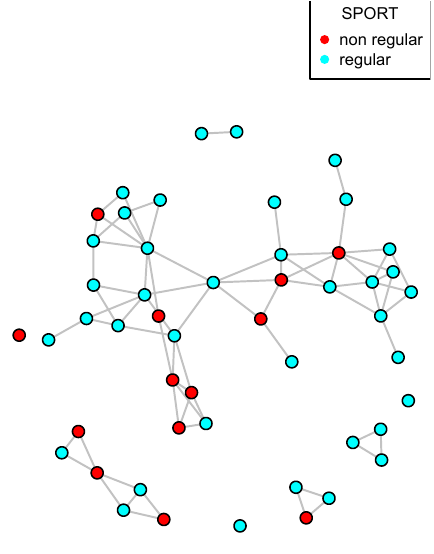}
\caption{50 girls from the Teenage Friends and Lifestyle Study dataset.}
\label{fig:teen_graphs}
\end{figure}

The adolescent friendship network were collected in the ``Teenage Friends and Lifestyle Study'' \citep{enlighten2700}. Friendship network data and substance use were recorded for a cohort of pupils in a secondary school in Glasgow (Scotland). 
Here we consider 3 actor covariates: drugs consumption (which was binarized in this example), sport activity, and smoking (Figure~\ref{fig:teen_graphs}).

In this example we focus on the transitivity effect expreessed by the geometrically weighted edgewise shared partner network statistic and the homophily effect of the drugs consumption (\code{nodematch("drugs")}) and its relationship to sport activity (\code{nodematch(c("sport","drugs")}) and smoking (\code{nodematch(c("smoke","drugs"))}).

The homophily effect modeled by \code{nodematch} counts the number of edges for which two nodes share the same covariate value. When multiple relationships are studied, the \code{nodematch} statistic counts only those on which all the covariate values match). More information about network statistics and their description can be found typing \code{?ergm.terms}.

We propose the following 4 competing models:
\begin{CodeChunk}
\begin{CodeInput}
R> m1 = y ~ edges + gwesp(log(2),fixed=TRUE) + nodematch(c("sport","drugs"))
R> m2 = y ~ edges + gwesp(log(2),fixed=TRUE) + nodematch(c("smoke","drugs"))
R> m3 = y ~ edges + gwesp(log(2),fixed=TRUE) + nodematch("drugs")
R> m4 = y ~ edges + gwesp(log(2),fixed=TRUE) 
\end{CodeInput}
\end{CodeChunk}
An important advantages of the Bayesian approach include easily interpreted measures of uncertainty through the use of prior knowledge. In this context, for example, it is known that the network graph is sparse meaning that the density effect (expressed by the \code{edge} statistic) is likely to be negative. For this reason we can include this prior information by setting the parameter value for the \code{edge} statistic equal to $-1$. 

\begin{figure}[htp]
\centering
\includegraphics{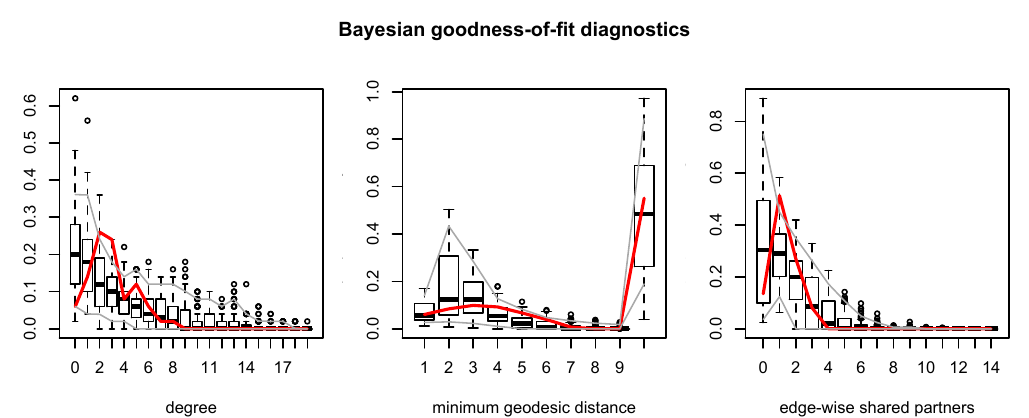}
\caption{Bayesian goodness-of-fit diagnostics.}
\label{fig:teen_bgof}
\end{figure}

We can also set up the prior variance/covariance structure. In this case we set the prior covariance matrix of each model to be a diagonal matrix with every entry equal to $5$.
\begin{CodeChunk}
\begin{CodeInput}
R> mean.priors <- list(c(-1,0,0),c(-1,0,0),c(-1,0,0),c(-1,0))
R> sigma <- 5 
R> sigma.priors <- list(diag(sigma,3),diag(sigma,3),diag(sigma,3),diag(sigma,2))
\end{CodeInput}
\end{CodeChunk}
As we have done above, we can use the \code{bergmS} function to perform Bayesian model selection and get an estimate of the Bayes Factors.
\begin{CodeChunk}
\begin{CodeInput}
R> mod.sel <- bergmS(c(m1,m2,m3,m4),
+                    iters=50000, 
+                    mean.priors=mean.priors,
+                    sigma.priors = sigma.priors,
+                    aux.iters=10000,
+                    main.iters=rep(1000,4),
+                    burn.ins=rep(100,4),
+                    gammas=rep(0.7,4))
R> best.mod <- bergmS.output(mod.sel,lag.max=200)
\end{CodeInput}
\begin{CodeOutput}
 BEST MODEL 
 ----------
Model 3: y ~ edges + gwesp(log(2), fixed = TRUE) + nodematch(c("drugs"))

 Posterior parameter estimate: 
                                       Post. mean:  Post. sd:
theta1 (edges)                          -4.5361571  0.3463194
theta2 (gwesp.fixed.0.693147180559945)   0.9865504  0.1248959
theta3 (nodematch.drugs)                 0.7787379  0.3217659

 Within-model acceptance rate: 0.11
 

Model 4: y ~ edges + gwesp(log(2), fixed = TRUE)

 Posterior parameter estimate: 
                                       Post. mean:  Post. sd:
theta1 (edges)                          -3.9527561  0.2040759
theta2 (gwesp.fixed.0.693147180559945)   1.0488695  0.1226025

 Within-model acceptance rate: 0.21 
 
BF_34 = 6.0360863656953


Model 2: y ~ edges + gwesp(log(2), fixed = TRUE) + nodematch(c("smoke", "drugs"))

 Posterior parameter estimate: 
                                       Post. mean:  Post. sd:
theta1 (edges)                          -4.1178225  0.3662144
theta2 (gwesp.fixed.0.693147180559945)   1.0396841  0.1585006
theta3 (nodematch.smoke.drugs)           0.2939651  0.1701672

 Within-model acceptance rate: 0.14 
 
BF_32 = 16.2310190824198


Model 1: y ~ edges + gwesp(log(2), fixed = TRUE) + nodematch(c("sport", "drugs"))

 Posterior parameter estimate: 
                                       Post. mean:  Post. sd:
theta1 (edges)                          -4.0539825  0.2115534
theta2 (gwesp.fixed.0.693147180559945)   1.0657870  0.1101090
theta3 (nodematch.sport.drugs)           0.1410472  0.1958435

 Within-model acceptance rate: 0.13 
 
BF_31 = 42.6648879402348

Between-model acceptance rate: 0.04
\end{CodeOutput}
\end{CodeChunk}
In the results above we have the posterior parameter estimates for two of the competing models with respective within-model acceptance rates and an estimate of the Bayes Factors for the comparison between the four models which makes clear that there is evidence that Model 4 is the best model of the set. This implies that the observed network is enhanced by the drugs consumption homophily effect.
\begin{CodeChunk}
\begin{CodeInput}
R> bgof(best.mod,
+       aux.iters=20000,
+       n.deg=20,
+       n.dist=10,
+       n.esp=15)
\end{CodeInput}
\end{CodeChunk}
In this example, the observed data appears to be a reasonable fit to the posterior distribution of the model selected (Model 3),
based on the goodness of fit geodesic and the shared partners statistics displayed in Figure~\ref{fig:teen_bgof}.

\begin{figure}[htp]
\centering
\includegraphics[height=7cm,keepaspectratio]{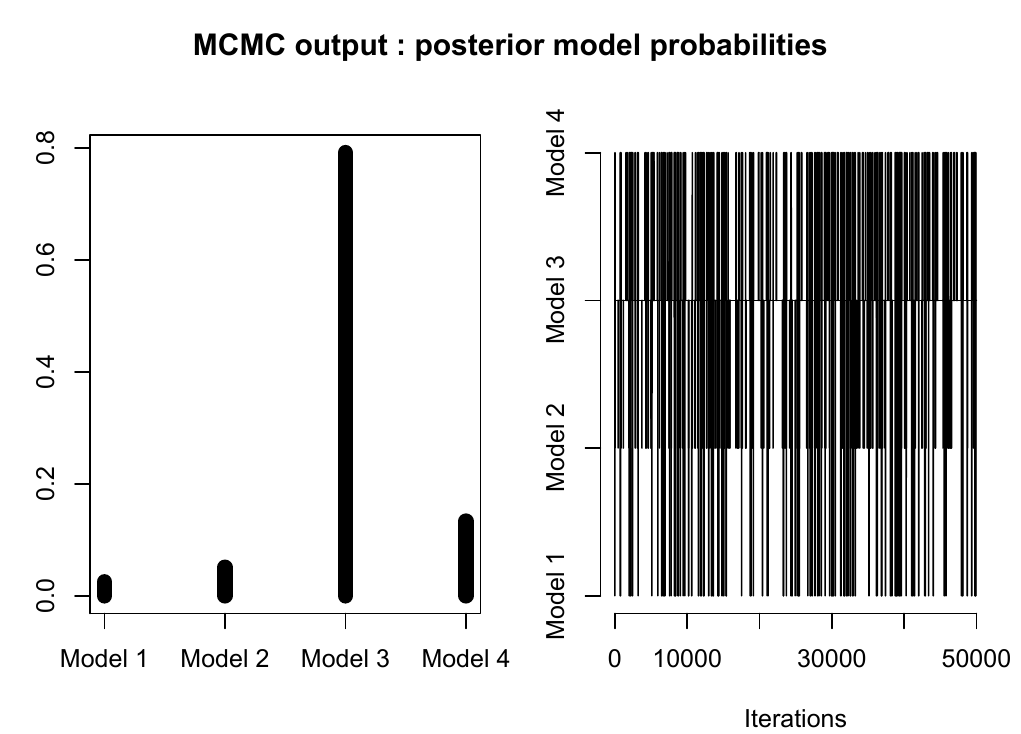}
\caption{MCMC diagnostics: posterior model probabilities.}
\label{fig:teen_modpost}
\end{figure}

\begin{figure}[htp]
\centering
\includegraphics[height=12cm,keepaspectratio]{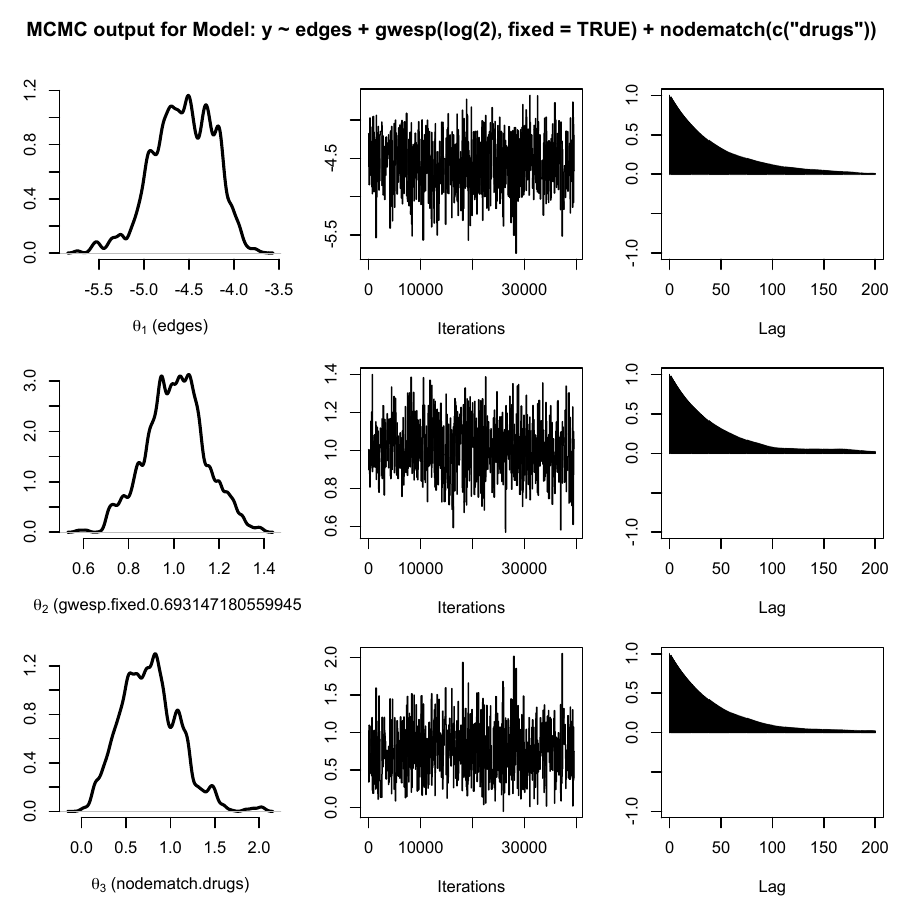}
\caption{MCMC diagnostics: posterior parameter probabilities.}
\label{fig:teen_parpost}
\end{figure}

From this results we can conclude that girls having the same level of drugs assumption tends to be friend. 
While girls with the same level of drugs assumptions and same level of smoking behaviour or sport activity do not seem to create a significant number of friendship connections. The transitivity effect is an important feature of the network graph but it is not sufficient to explain the  complexity of the observed network data.

\section[Discussion]{Discussion}

The software package \pkg{Bergm} aims to help researchers and practitioners in two ways. Firstly, it is currently the only package 
for \proglang{R} that provides a simple and complete range of tools for conducting Bayesian analysis for exponential random graph models. 
Secondly, \pkg{Bergm} makes available a platform that can be easily customised, extended, and adapted to address different 
requirements.

The software package is under continual development and it is far from finished. The main limitation of the software is its 
computational cost which makes it unsuitable for managing network graphs larger than hundreds of nodes, however it is perfectly
suited for networks involving up to a hundred nodes. An important improvement in terms of computational time and efficiency 
will be done by turning some of the \proglang{R} functions into \proglang{C} functions integrated with the \pkg{ergm} package. We expect that will yield further reductions in computational run time. 

Future versions of the \pkg{Bergm} package will address several issues including Bayesian analysis of Curved Exponential Random 
Graph Models \citep{hun:han06} and exponential random graph models with missing data \citep{kos:rob:pat10}.

\newpage
\bibliography{myref}

\newpage
\section*{A. \proglang{R} code for loading and plotting the network data}
\label{appendix}

Here we give the code to load the datasets presented in this paper and to plot the network graphs displayed in Figure~\ref{fig:graph}. The main functions to load and plot network data are \code{network} and \code{plot.network} respectively. They are both included in the \pkg{network} package which is one of the depencencies of \pkg{Bergm} and it is automatically loaded by typing the \code{library("Bergm")} command (see Section~\ref{sec:getbergm}). We include the \code{set.seed} statement in order to produce exactly the same graphs of Figure~\ref{fig:graph} and \ref{fig:teen_graphs}.
The datasets used in this paper are available in the \pkg{statnet} package. The nodal attributes for the Kapferer dataset and the Teenage Friends and Lifestyle Study' social network dataset can be downloaded from the Siena datasets repository \url{http://www.stats.ox.ac.uk/~snijders/siena/}.
\begin{CodeChunk}
\begin{CodeInput}
R> install.packages("statnet")  
R> library("statnet")
R> install.packages("Bergm")  
R> library("Bergm")
\end{CodeInput}
\end{CodeChunk}

The Kapferer Taylor Shop network and nodal attribute data can be loaded into \proglang{R} by typing:
\begin{CodeChunk}
\begin{CodeInput}
R> data(kapferer)
R> y <- kapferer
R> x <- read.table("~/kapfa_stat.dat")
R> y %v% 'job' <- x
\end{CodeInput}
\end{CodeChunk}
The last command is used to attache the nodal covariate ``job'' represented by the object \code{x} to the network object \code{y}.

To plot the network graph as in Figure~\ref{fig:graph1} we used the following code:
\begin{CodeChunk}
\begin{CodeInput}
R> CC <- colors()[c(24,135,53,142,258,28,551,119)]
R> set.seed(20)
R> par(oma=rep(0,4),mar=rep(0,4))
R> plot(y,
+       vertex.col=CC[x[,1]],
+       edge.col=colors()[c(229)],
+       vertex.cex=1.5,
+       usearrows=TRUE)
R> legend(8,8, 
+         legend=seq(1,8), 
+         col=CC,
+         yjust=0,
+         pch=19) 
\end{CodeInput}
\end{CodeChunk}

The \code{legend} function creates a legend showing the colors associated to the levels of the ``job'' nodal attribute.  For more information about this function type \code{?legend}.

The following code was used to load the Zachary Karate Club network and to plot it as in Figure~\ref{fig:graph}: 
\begin{CodeChunk}
\begin{CodeInput}
R> data(zach)
R> y <- zach
R> par(oma=rep(0,4),mar=rep(0,4))
R> plot(y,
+       vertex.col=colors()[123],
+       edge.col=colors()[c(229)],
+       vertex.cex=1.5)
\end{CodeInput}
\end{CodeChunk}

The following code was used to load the Teenage Friends and Lifestyle Study' social network and to plot it as in Figure~\ref{fig:teen_graphs}: 
\begin{CodeChunk}
\begin{CodeInput}
R> y <- read.table("~/s50-network1.dat",sep=''))
R> y <- network(as.matrix(y),matrix.type="adjacency",directed=FALSE)
R> x1 <- read.table("~/s50_data/s50-sport.dat")
R> x2 <- read.table("~/s50_data/s50-smoke.dat")
R> x4 <- read.table("~/s50_data/s50-drugs.dat")
R> y %v% 'sport' <- x1[,1]
R> y %v% 'smoke' <- x2[,1]
R> y %v% 'alcohol' <- x3[,1]
R> x4[,1][x4[,1]<3]=1
R> x4[,1][x4[,1]>2]=2
R> y %v% 'drugs' <- x4[,1] 
R> CC <- rainbow(2)
R> set.seed(20)
R> par(mfrow=c(1,3),oma=rep(0,4),mar=rep(0,4))
R> plot(y,
+       vertex.col=CC[x1[,1]],
+       edge.col=colors()[c(229)],
+       vertex.cex=1.5,
+       usearrows=TRUE)
R> legend("topright", 
+         legend=c("non regular","regular"), 
+         col=CC,
+         yjust=0,
+         pch=19,
+         title="SPORT") 
R> CC <- rainbow(3)
R> set.seed(20)
R> plot(y,
+       vertex.col=CC[x2[,1]],
+       edge.col=colors()[c(229)],
+       vertex.cex=1.5,
+       usearrows=TRUE)
R> legend("topright", 
+         legend=c("non","occasional","regular"), 
+         col=CC,
+         yjust=0,
+         pch=19,
+        title="SMOKE") 
R> CC <- rainbow(2)
R> set.seed(20)
R> plot(y,
+       vertex.col=CC[x4[,1]],
+       edge.col=colors()[c(229)],
+       vertex.cex=1.5,
+       usearrows=TRUE)
R> legend("topright", 
+         legend=c("non - once or twice a year","once a month - once a week"),
+         col=CC,
+         yjust=0,
+         pch=19,
+         title="DRUGS") 
\end{CodeInput}
\end{CodeChunk}
More details about the commands used in this section can be found in the help files by typing \code{?network} and \code{?plot.network}. More information about the \pkg{network} package can be found in \cite{but08}.

\end{document}